# The solution to an unresolved problem: newly synthesised nanocollagen for the preservation of leather.

Marina Bicchieri[a]*, Federica Valentini[b,c], Francesca Pascalicchio[d], Maria Luisa Riccardi[e], Piero Colaizzi[f], Camilla Del Re[g], Maurizio Talamo[c]

- [a.] Istituto centrale restauro e conservazione patrimonio archivistico e librario, Chemistry Department, Via Milano 76, 00184 Roma, Italy, marina.bicchieri@beniculturali.it
- [b.] Università Tor Vergata, Chemistry Department, Via della Ricerca Scientifica 1, 00133 Roma, Italy, federica.valentini@uniroma2.it
- [c.] INUIT Università Tor Vergata, Via dell'Archiginnasio snc, 00133 Roma, Italy, maurizio.talamo@inuitroma2.it
- [d.] Istituto centrale restauro e conservazione patrimonio archivistico e librario, Technology Department, Via Milano 76, 00184 Roma, Italy, francesca.pascalicchio@beniculturali.it
- [e.] Istituto centrale restauro e conservazione patrimonio archivistico e librario, Restoration Department, Via Milano 76, 00184 Roma, Italy, marialuisa.riccardi@beniculturali.it
- [f.] Istituto centrale restauro e conservazione patrimonio archivistico e librario, Biology Department, Via Milano 76, 00184 Roma, Italy, piero.colaizzi@beniculturali.it
- [g.] Istituto centrale restauro e conservazione patrimonio archivistico e librario, SAF, Via Milano 76, 00184 Roma, Italy, delrecamilla@gmail.com

*Corresponding author: Marina Bicchieri

**Abstract**

A widespread problem in libraries is related to the preservation of book covers in leather that are often torn, powdery and abraded. The same problem is encountered in the conservation of leather goods. Until now a satisfactory solution to contrast the leather deterioration had not been found and the applied conservation methods offered only temporary solutions, without guaranteeing a real and durable effectiveness. At the Istituto centrale restauro e conservazione patrimonio archivistico e librario (Icrcpal) it was decided to research more durable results and to apply nanocollagen solutions to the leather. A new synthesis of nanocollagen was performed in collaboration with Università Tor Vergata, and Fondazione INUIT and the newly synthesised nanocollagen was characterised by different spectroscopic and imaging techniques, then applied to laboratory samples and, at the end of the research, it was used in the restoration of the leather cover of a 18[th] book. All the measurements performed on the tested leathers did not show any colour change after nanocollagen application, an increase of all mechanical characteristics and, of paramount importance, an increase in the shrinkage temperature of the leather with a partial reconstitution of its lost elasticity and flexibility.

## 1. Introduction

The preservation of torn, powdery, lacunose, worn, abraded, weak and friable book-covers in leather is a real problem for library conservators. These kinds of deterioration are linked to the ageing, the usage and manipulation of the books, the interactions with pollutants, but they are also connected with the products used in the leather manufacture or in the finishing treatments applied for special or decorative purposes.

Moreover, the choice of the materials for the covers was very often more related to their price than to their durability and permanence. Sheep leather, a less durable and stable material in respect to calf or goat leathers, is one of the most widespread materials in the history of bookbinding, especially from 17th century, due to its lower market price.

When a cover book is severely deteriorated, a normal manipulation of the book is almost impossible, without causing the detachment of small fragments or "dust" of leather from the binding, and a restoring treatment is needed.

At present no satisfactory and durable solutions to contrast the leather deterioration have been found, but it is possible to apply environmental control strategies and preventive conservation practices [1]. If environmental control is always advisable, the preventive conservation is rather applicable to museum objects than to artifacts that should be consulted and used, such as books.

Since the 70s of the last century hydroxypropylcellulose [2] was frequently used for the consolidation of leather. More recently a mixture of waxes and acrylic resin (SC 6000) was proposed and used alone or in mixture with hydroxypropylcellulose [3].

None of these products had the capability to penetrate into the bulk of the leather that after the treatment often showed color changes and increased brittleness.

The challenge has always been the recovery of flexibility and stability of the degraded leather, without altering the appearance and the equilibrium of the internal fats and moisture.

## 2. Aim

The Istituto centrale restauro e conservazione patrimonio archivistico e librario (Icrcpal) decided, in collaboration with the Fondazione INUIT, to approach the problem in a different way by using nanomaterials expressly designed for the conservation of leather. Nanomaterials can, in fact, penetrate into the bulk of the treated material, offering a deeper consolidation effect. The idea was to treat the leather with its same principal component, the collagen, synthesised at nanoscale dimension and, after a series of laboratory experiments, to apply it in a real restoration case study.

A promising preliminary investigation was performed in 2014 [4, 5], but not implemented. Recently a new electrochemical synthesis was optimised and patented (Patent N. 102016000096336, 2016) obtaining a nanocollagen with enhanced dispersibility in different working media and long-term stability of the colloidal phase dispersion (over 1 year, at ambient temperature, without precipitation of a solid phase).

To assess the final application procedures, the effect on leather of the newly synthesised nanocollagen, soluble both in isopropyl alcohol and in water, was studied by optical measurements, chemical and mechanical tests. Moreover different nanocollagen concentrations and solvents were tested on laboratory samples, to determine the minimal amount of product to be applied to the leather with the best consolidation results.

After a preliminary study, an original book was chosen for the real application.

## 3. Experimental

*3.1. Materials*

*3.1.1. Reagents for the synthesis of Nanocollagen.*
Bovine Collagen Type I (Sigma-Aldrich) was used as molecular precursor for the synthesis of nanocollagen, performed in acetate buffer aqueous solution 0.1 M, pH 4.7, (Sigma-Aldrich). Alumina/$Al_2O_3$ tracked etched template membranes (Whatman® Anodisc Inorganic Membranes) were used (membrane diameter 30 mm, pores diameter 200 nm, pores length 100 μm, pores density $1 \times 10^{12}$ pores/$cm^2$), as well as $HNO_3$ and $NaOH$ (Sigma-Aldrich, analytical grade).

*3.1.2. Laboratory leathers*
In the preliminary tests, the vegetable tanned calfskin leather (mean thickness 1.6 mm) of a 18th century cover, contemporary with the original volume, was used. For further tests a calfskin grain split leather (chrome tanned, coloured with soluble aniline dyes, mean thickness 0.66 mm) was employed, because it showed mechanical characteristics similar to those of the original cover.

*3.1.3. Original cover*
The *Estro Poetico Armonico* by Benedetto Marcello (Mus. 243, 18th century, Biblioteca Casanatense, Roma) was a perfect case study, presenting all the damages described in the introduction. It belongs to a series of five books with the same binding (first edition *in folio*, Venezia, 1724-26), thus allowing for a comparison with other original specimens, after the treatment. The cover of Mus. 243 (mean thickness 0.60 mm), a vegetable sheepskin leather, tanned with hydrolysable tannins, after the manufacture was mottled with an acidic solution.

*3.1.4. Tanning detection*
The tanning was detected by specific spot tests: ferric chloride for vegetable tannins, rhodanine for hydrolysable tannins, acid butanol for condensed tannins, alizarin sulphonate for aluminium detection [6, 7, 8].

### 3.2. Methods

*3.2.1. Electrochemical synthesis of nanocollagen.*
ChronoAmperometry was the electrochemical techniques applied for the synthesis of nanocollagen. The optimised parameters were patented in 2016 (Patent N. 102016000096336, 2016) and are briefly discussed below.
The tropocollagen precursor was used at concentration 1 mM in 0.1 M acetate buffer solution at pH 4.7. To assemble the Alumina Template Working Electrodes (ATWEs), it was necessary to make the $Al_2O_3$ membrane conductive. An Ag layer (20 nm thickness) was then deposited by sputtering for 2 min at 2 mA. During the electrochemical synthesis, a constant and controlled working potential value of -1.0 V/versus $Ag/AgCl/Cl^-$ reference electrode was applied by ChronoAmperometry with deposition time 3600 s under $N_2$ at flow rate of 0.3 $cm^3$/min. During the ChronoAmperometry deposition, the electrolysis solution was magnetically stirred, at ambient temperature. After the electrochemical synthesis of nanocollagen, the silver conductive layer was dissolved in concentrated $HNO_3$ and the alumina template membrane was removed with concentrated NaOH solution unable to dissolve the nanocollagen that was rinsed in water until neutrality.

*3.2.2. TEM (Transmission electron Microscopy)*
TEM Philips Electron Optics 301 was employed for the morphological study of nanocollagen samples prepared by coating Cu grids (φ=3mm), by deep coating in 0.7 mg/ml of nanocollagen dispersion. After immersion, the coated Cu grids were dried under Wood lamp.

*3.2.3. SEM (Scanning Electron Microscopy)*
For the morphological characterisation of nanocollagen a FE-SEM/EDX, LEO 1550 equipped with a sputter coater (Edwards Scan Coat K550X) was used. A volume of 5 μL of the nanocollagen dispersion was deposited on Si(111), allowing the solvent to evaporate at room temperature, and then fixed on aluminium stub with carbon tape. The samples were then coated with Au layer (thickness 10-20 nm), deposited by sputtering for 2 min at 25 mA.
For the studies on the application of nanocollagen to leather, the SEM analyses were performed with a Carl-Zeiss EVO 50 instrument equipped with both a detector for electron-backscattered diffraction (BSD) and for Secondary Electron scanning in Variable Pressure mode (VPSE). SEM observations were performed at 20 kV accelerating voltage with a tungsten filament. All the samples were mounted on Al stubs and observed at different magnification ranging from 500X to 3000X, following a fixed measurement grid, before and after the treatment with nanocollagen, which was applied on the sample directly in the SEM chamber, in order to repeat the observations after the consolidation, exactly at the same point observed before the treatment. It was possible to perform the analyses also on some original fragments spontaneously detached from the binding and no more repositionable.

*3.2.4. FT-IR (Fourier Transform-Infrared Spectroscopy)*
IR spectra for the nanocollagen structural characterisation were recorded by a Perkin-Elmer Spectrum One FT-IR spectrometer from KBr pellets in $N_2$ environment.

*3.2.5. Colour coordinates*
A Minolta Chroma Meter CR22 colorimeter was used in the CIE L* a* b* space, averaging 3 measurement for each analysed point. Delta E before and after the treatments was calculated in agreement with ISO/CIE 11664-6:2014(E) [9].

*3.2.6. Tear resistance*
For the measurements a Buchel Van Der Korput Tearing Tester (Elmendorf Type) was used. The number of samples (h: 60 mm, w: 50 mm) varied as a function of the amount of leather that could be subjected to the tests as will be evidenced in the Results section. Tear resistance was measured in accordance with T 414 om-12 method [10] that was used both to measure the resistance to the formation of a tear (tear initiation) and the resistance to the expansion of a tear (tear propagation).
It was decided to extend to leather a method normally used for paper, because the standardised methods for leather are conceived for commercial products with higher mechanical characteristics in respect to those of the samples chosen to simulate the original cover.

*3.2.7. Tensile force*
For the measurements an Instron Tensile Tester Model 1026 was used with: load cell 0.49 kN at full scale, load range adjustable with a scale selector, crosshead speed 100 mm/min. The dimensions of the samples were: h 200 mm, w 15 mm; useful length 150 mm.

*3.2.8. Bending resistance (Stiffness)*
For the measurement a Lorentzen & Wettre, Type 10-1 Bending Tester was used.
The samples had dimensions 70 mm (h) and 38 mm (w) and the measurements were carried out with a bending angle φ 30° and bending length 25 mm. The number of samples varied as a function of the amount of leather that could be subjected to the tests as will be evidenced in the Results section.

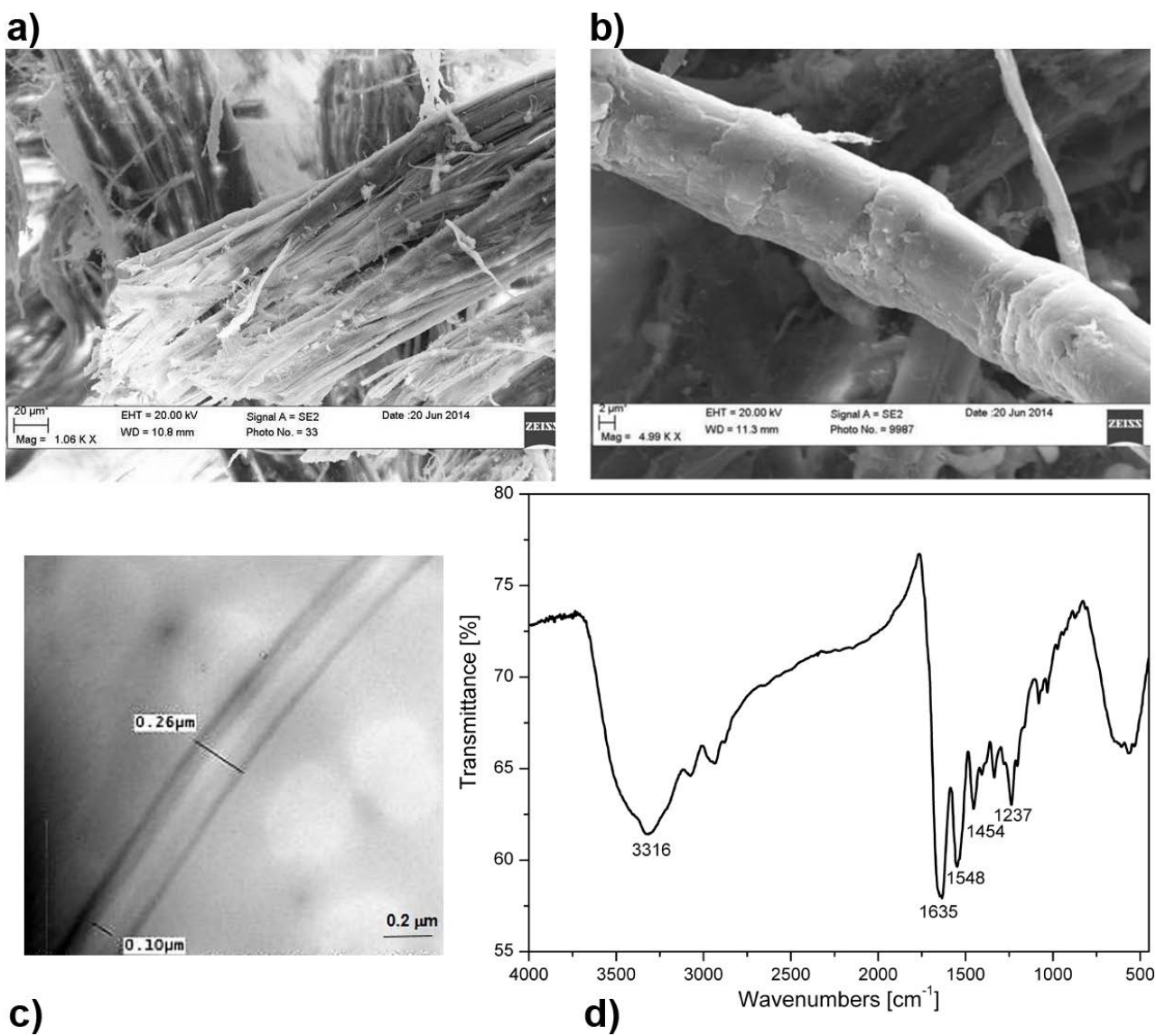

**Fig. 1.** Characterisation study of the new nanocollagen. (a): SEM micrograph of a typical bundle of tropocollagen, molecular precursor for the synthesis of collagen nanotubes. (b): Single fibre from tropocollagen bundles, before the electrochemical template synthesis route. (c): TEM of the new synthesised nanocollagen; (d): FTIR spectrum of the new nanocollagen samples. Wavelength of the characteristic collagen peaks are reported.

*3.2.9. pH and pH difference*
pH of the original cover was measured by a portable Crison pH-25 equipped with a flat membrane glass electrode 5027, after extraction of 0.25 g of leather in 5 ml water. pH difference was measured by diluting 10 times the solution used for pH measurements (cold extraction standard test ISO 4045:2008 [11]).

*3.2.10. Shrinkage temperature*
Very small amount of fibres were removed from the flesh side of the leather, wetted with distilled water for 10 min on a single concave microscope slide. The fibres were then separated with a needle, covered with water and a coverslip. The microscope slide was placed on the hot table Mettler FP82 Hot Stage, thermostatically controlled through a Mettler FP90 Central Processor, and heated at a rate of 2°C/min. The samples were observed under a Leica DMLP microscope in transmitted light, at 50X magnification and video recorded by Leica MC 170 HD. Video recording allowed for the accurate verification of the temperature at which shrinkage events occurs. The method is described in ISO 3380:2015 [12] but had been modified in order to be applied in conservation field, where only small amount of material can be examined [13, 14]. In this work the modified method was applied.

*3.2.11. Microscopy*
To determine the animal species, the analysis of the hairs follicular patterns of the leathers was performed. The different leathers were observed with a Leica Macroscope M420, using cold incident light produced by optical fibres.

**4. Results and Discussion**
The experimental work has been divided in different steps that will be discussed in separate paragraphs.

*4.1. Characterisation of nanocollagen*
Fig. 1 shows the SEM and TEM morphological characterisation of the tropocollagen precursor before its modification into nanocollagen, and the nanocollagen final product. Fig. 1a, shows a typical bundle of tropocollagen fibres, one of which is reported at higher magnification in Fig. 1b to evidence its typical cylindrical shape. After the electrochemical tracked etched membrane template synthesis approach, the cylindrical fibre exhibits nanometer dimensions: inner diameter 100 nm, external diameter 260 nm (Fig. 1c, TEM micrograph). FTIR spectrum (Fig. 1d) confirms that the chemical structure of the tropocollagen precursor (Collagen Type I) is preserved after the synthesis. In particular, the band centred around 1635 cm$^{-1}$, is the typical absorption of Amide I [15, 16] due to the stretching vibration of the peptide carbonyl group (-C(=O)-). The spectral assignments are reported in Table S1, ESI section.

*4.2. Application on leather - preliminary investigation*
To evaluate the effect of nanocollagen on the leather, mechanical, optical and microscopic tests (tensile force, stiffness, colour coordinates, SEM imaging) were performed on a 18$^{th}$ century and no more usable leather cover book, chosen to simulate the original document. The cover was divided into 4 samples sets. Three sets were treated with different nanocollagen solution (Table 1), one was left untreated, and used as control.

*Table 1*
*Solutions used in the preliminary investigation on a 18$^{th}$ century leather cover*

| Solvent(s) | nanocollagen concentration |
|---|---|
| isopropyl alcohol/water 70/30 (V/V) | 0.7 mg/ml. |
| isopropyl alcohol/water 70/30 (V/V) | 1 mg/ml |
| water | 1 mg/ml |

The different nanocollagen concentration(s) and solvent(s) were chosen to determine both the best solvent and the more suitable concentration to be applied in the consolidation treatments. Due to the reduced amount of leather it was only possible to obtain 16 samples for the tensile force measurements and 6 samples for stiffness. The results are reported in Tables 2 and 3.
Concerning the tensile stress measurements (Table 2), the unevenness of the ancient cover, caused by the widespread walkways across the surface, gave randomly distributed values. For this reason, only the recorded values are reported, without statistical treatment, as well as for binding measurements. The unique result that can be inferred is that the samples treated with nanocollagen in hydroalcoholic solution (0.7 mg / ml) presented a better homogeneity regards the tensile stress.

*Table 2*
*Tensile force and breaking time of samples from a 18$^{th}$ century leather cover, before and after treatment with nanocollagen*

| Sample | Force (N) | Breaking time (s) |
|---|---|---|
| Non-treated samples | | |
| 1 | 78.5 | 12.2 |
| 2 | 145.1 | 14.5 |
| 3 | 35.3 | 4.8 |
| 4 | 54.9 | 6 |
| Samples treated with nanocollagen 0.7 mg/ml in isopropyl alcohol/water 70/30 (V/V) solution | | |
| 5 | 59.8 | 8.3 |
| 6 | 59.8 | 8 |
| 7 | 59.8 | 7 |
| 8 | 113.8 | 11 |
| Samples treated with nanocollagen 1 mg/ml in isopropyl alcohol/water 70/30 (V/V) solution | | |
| 9 | 51.0 | 6 |
| 10 | 19.6 | 2.5 |
| 11 | 66.7 | 8.2 |
| 12 | 53.0 | 7.1 |
| Samples treated with nanocollagen 1 mg/ml in water solution | | |
| 13 | 41.2 | 4.2 |
| 14 | 58.6 | 6 |
| 15 | 80.4 | 8.2 |
| 16 | 127.5 | 13 |

The binding measurements (Table 3) did not show a great variation in stiffness before and after all the treatments. The slight increase noticed after the nanoparticles application is ascribable to the addition of nanocollagen fibres to the leather, which filled the inter-fibres cavities. This result is particularly positive because consolidation treatments should not cause an excessive increase in the material rigidity that could lead to the breaking of the leather under stress and manipulation.
Table 4 contains the averaged values (4 samples for each nanocollagen solution) of colour coordinate and Delta E, before and after application of the different solutions of nanocollagen. None of the treatments induced noticeable colour variations, except a light decrease of luminosity. Also for colorimetric measures better results with lower optical impact were obtained when the hydroalcoholic solution 0.7 mg/ml was used.
The positive effect of nanocollagen treatment with all of the applied solutions was clearly highlighted by SEM images, as shown in Fig. 2. After the application, fibres appeared less tangled than before and the leather surface seemed smoother.

*Table 3*
*Stiffness of samples from a 18$^{th}$ century leather cover, before and after treatment with nanocollagen*

| Nanocollagen 0.7 mg/ml in isopropyl alcohol/water 70/30 (V/V) | | |
|---|---|---|
| Sample | Before treatment (mN) | After treatment (mN) |
| 1 | 877 | 950 |
| 2 | 797 | 798 |
| Nanocollagen 1 mg/ml in isopropyl alcohol/water 70/30 (V/V) | | |
| Sample | Before treatment (mN) | After treatment (mN) |
| 3 | 1127 | 1187 |
| 4 | 845 | 939 |
| Nanocollagen 1 mg/ml in water | | |
| Sample | Before treatment (mN) | After treatment (mN) |
| 5 | 1197 | 1212 |
| 6 | 917 | 1216 |

*Table 4*
*Colour coordinates of samples from a 18th century leather cover, before and after treatment with nanocollagen. Average on 4 samples for each treatment*

| Nanocollagen 0.7 mg/ml in isopropyl alcohol/water 70/30 (V/V) | | | |
|---|---|---|---|
| Coordinate | Before treatment | After treatment | Delta E CIE 2000 |
| L* | 28.32 ± 0.78 | 28.01 ± 0.79 | |
| a* | 14.31 ± 0.66 | 14.03 ± 0.75 | 0.27 |
| b* | 12.66 ± 1.33 | 12.62 ± 0.41 | |
| Nanocollagen 1 mg/ml in isopropyl alcohol/water 70/30 (V/V) | | | |
| Coordinate | Before treatment | After treatment | Delta E CIE 2000 |
| L* | 32.64 ± 4.05 | 30.19 ± 4.28 | |
| a* | 13.53 ± 1.08 | 14.32 ± .16 | 1.92 |
| b* | 14.63 ± 2.93 | 14.56 ± 3.42 | |
| Nanocollagen 1 mg/ml in water | | | |
| Coordinate | Before treatment | After treatment | Delta E CIE 2000 |
| L* | 29.64 ± 1.31 | 29.30 ± 1.62 | |
| a* | 14.98 ± 0.62 | 15.38 ± 0.71 | 0.39 |
| b* | 12.86 ± 0.84 | 13.09 ± 0.96 | |

The SEM analyses showed the reconstructing effect due to the nanoparticles application: the cover margin that was disordered and spread apart prior to the treatment (Fig. 2 bottom left) appeared rebuilt in its integrity (Fig. 2 bottom right) after the treatment.

*4.3. Application on leather - evaluation of the effects of nanocollagen at the working concentration*
The encouraging results obtained in the preliminary investigation gave the possibility both to choose the optimal concentration (0.7 mg/ml in hydroalcoholic solution) and to perform a new series of tests on a more homogeneous set of samples. In this second phase of the research a modern industrial split leather was used. This kind of leather was very thin, almost lacking in the flesh layer and presented mechanical characteristics very similar to those of the original leather needing restoration. In this phase of the work, tear resistance tests were added to the other mechanical tests to verify the ability of the nanocollagen solution to reinforce the leather, in particular in regards to its capability to resist to tearing stress, an important parameter for the manipulation of real objects.
Table 5 contains the results of the mechanical and optical tests. Only the averaged values are reported, as well as the number of samples subjected to the tests.
As can be seen in Table 5, there is a sensible increase in stiffness after the treatment, indicating an increment of the resistance of the material. Fortunately, the recorded absolute values do not correspond to a rigid and breakable leather and the increase can be regarded as a positive effect for the stability of the treated leather.
Very positive results were obtained from the tensile stress measurements (Table 5). Apart the increase in the force necessary to induce the breaking of the sample, the most positive effect is the increment in the elongation parameter that indicates an augmented elasticity of the treated leather, which can better resist to the mechanical stresses induced by manipulation.
Concerning tear resistance measures, two different kinds of tests were performed: tear propagation and tear initiation. The first measure simulates a very common problem: the existence of tears in a real object that can prosecute when the object is manipulated. The second kind of measure gives information on the resistance of the margin of an object, a cover in this case, both during the manufacture/restoration and the usage.
The tear propagation was measured on single sample and on two coupled samples.
The results (Table 5) show a very important increment of the tear resistance after the treatment with nanocollagen, in particular regards the tear initiation. This increase is very positive for the future manipulation of the original restored artefact.

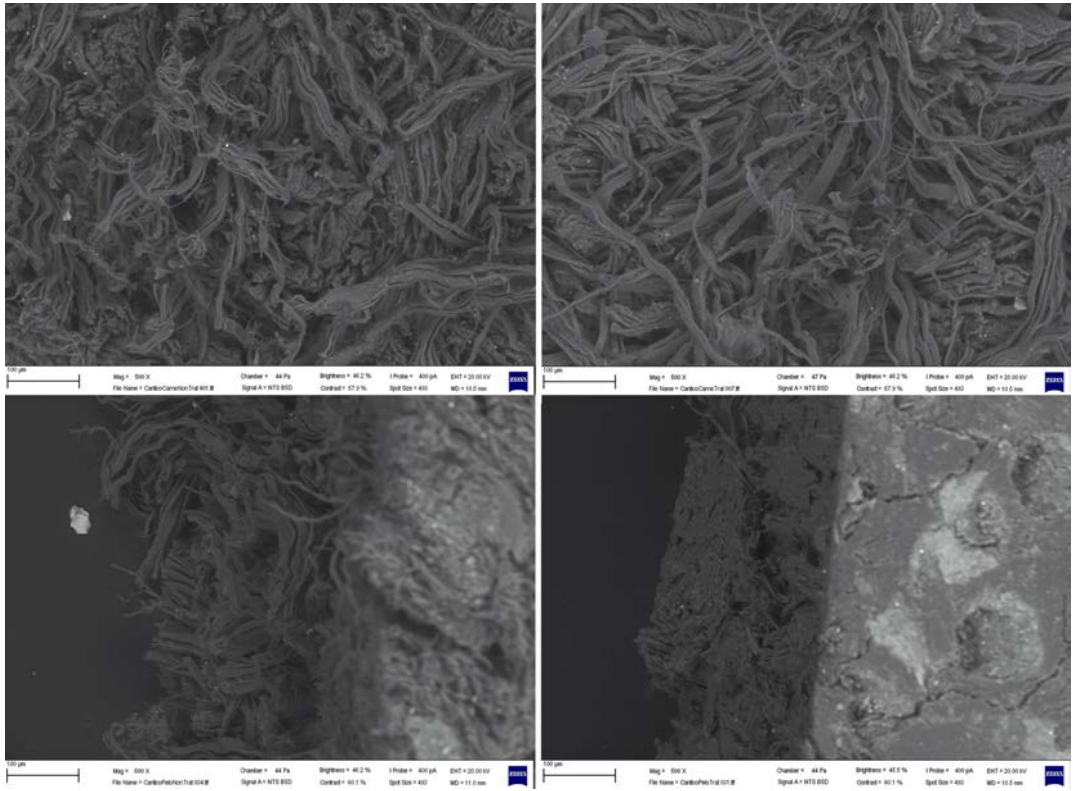

**Fig. 2.** Ancient cover, 18th century. Left side: before treatment with aqueous nanocollagen 1 mg/ml. Right side: after treatment. Top: flesh side of the leather; bottom: grain side. Magnification 500X.

*Table 5*
*Mechanical and optical characteristics of the split-leather before and after the treatment with nanocollagen 0.7 mg/ml in isopropyl alcohol/water 70/30 (V/V)*

| Stiffness (mN) | | | | | |
|---|---|---|---|---|---|
| Before treatment | | | After treatment | | |
| 7 ± 2 | | | 19 ± 8 | | |
| Tensile stress | | | | | |
| Before treatment | | | After treatment | | |
| Force (N) | Breaking time (s) | Elongation (mm) | Force (N) | Breaking time (s) | Elongation (mm) |
| 38.9 ± 6.2 | 41 ± 2 | 23 ± 2 | 43.0 ± 5.1 | 48 ± 5 | 27 ± 2 |
| Tear propagation (mN) | | | | | |
| Before treatment | | | After treatment | | |
| 2126 ± 53 (two coupled samples) | | | 2243 ± 87 (two coupled samples) | | |
| 3567 ± 90 (single sample) | | | 3782 ± 185 (single sample) | | |
| Tear initiation (mN) | | | | | |
| Before treatment | | | After treatment | | |
| 4081 ± 80 (single sample) | | | 6670 ± 92 (single sample) | | |
| Colour Coordinates | | | | | |
| Coordinate | Before treatment | | After treatment | | Delta E CIE 2000 |
| L+ | 28.38 ± 0.26 | | 31.91 ± 0.37 | | |
| a* | 16.77 ± 0.25 | | 15.08 ± 0.22 | | 3.26 |
| b* | 22.76 ± 1.96 | | 19.45 ± 0.47 | | |

Stiffness: 48 samples. Tensile stress: 30 samples. Tear propagation: 24 samples. Tear initiation: 10 samples. Colorimetric measurements: 24 samples.

Values of colour coordinates and Delta E (Table 5) show that in the case of the split leather the final colour difference is greater than those obtained on the ancient 18[th] century cover, probably as a consequence of the dye used for colouring the leather: vegetable in the ancient cover and aniline for the modern split leather. The main difference is related to luminosity that increases after the treatment.

SEM images of the split leather showed the same improvement noticed for the ancient 18th century cover.

*4.4. Application on leather - final work on the original volume Mus. 243*

All the very positive results, previously reported and commented, allowed to apply the nanocollagen solution (0.7 mg/ml in isopropyl alcohol/water 70/30 (V/V)) to the original cover, after performing SEM imaging analysis on non-repositionable fragments. In Fig 3 some images are reported and the arrows highlight peculiar effects such as the relaxation of the fibres (Fig 3 top) or the formation of bonds between fibres (Fig. 3 middle) or the filling effect of nanocollagen (Fig. 3 bottom).

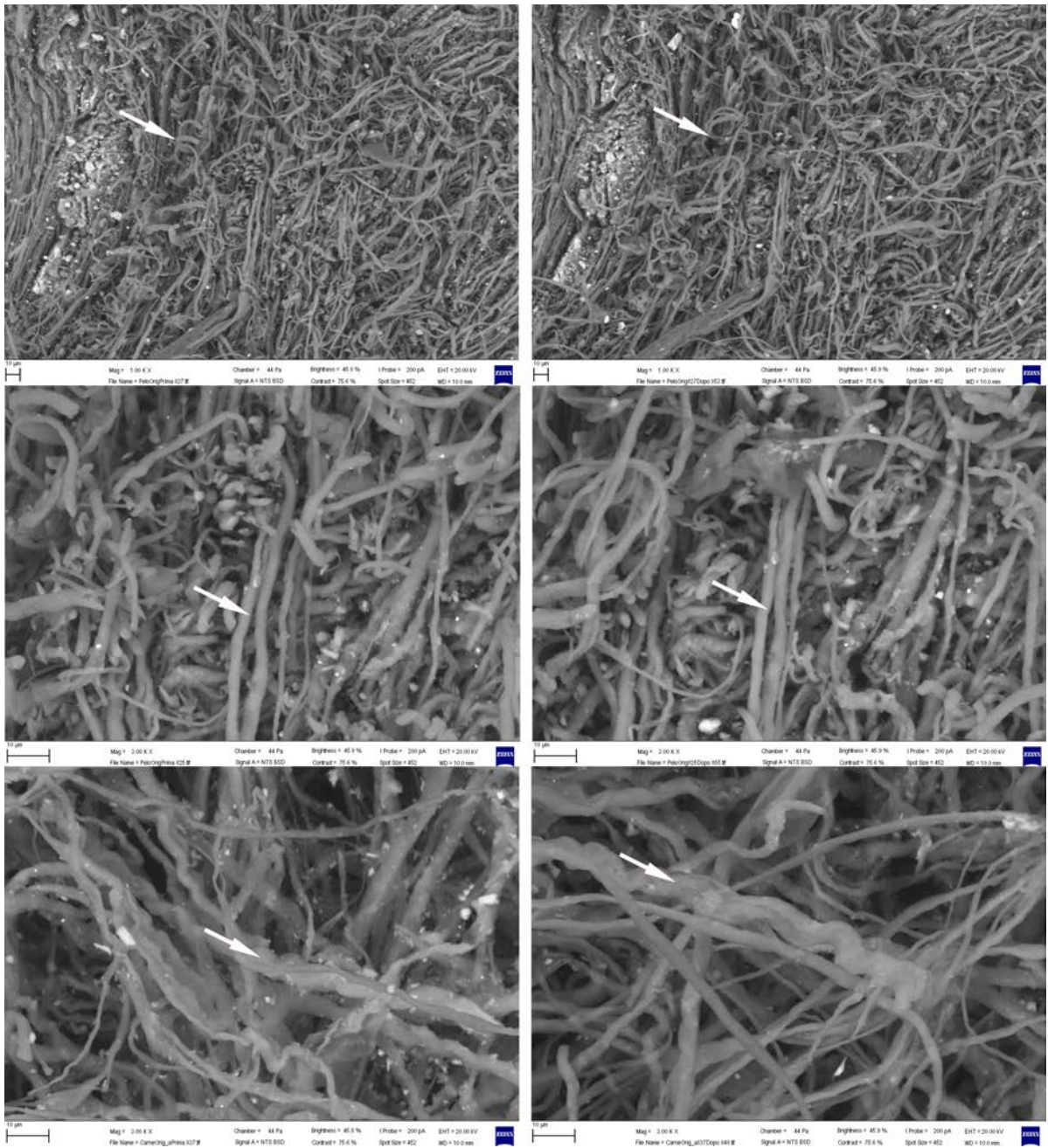

**Fig. 3.** Mus. 243. Left side: before treatment with nanocollagen 0.7 mg/ml in isopropyl alcohol/water 70/30 (V/V). Right side: after treatment. Top: flesh side of the leather, magnification 1000X; middle: grain side 3000X, bottom: grain side 3000X. The arrows highlight some peculiar features and the effect of nanocollagen.

In addition to mechanical and optical tests, measurements of pH, pH difference and shrinkage temperature were performed.

Mus. 243 had average pH 4.3. It is well known that in the 3-4 pH range leather is unharmed by the acidity [17], and the kind or amount of tannin used in manufacturing has little effect on pH, which can also be influenced by the interactions with external factors, such as light, ionising radiations and pollutants. More or less weak acids can therefore be formed into the leather and their presence can be revealed by pH difference measures, i.e. the difference between the standardised pH measurement and the pH of the same solution diluted ten times [11]. This difference is ranging from 0 to 1, and a measure between 0.7 and 1 denotes the presence of a certain amount of strong free acids, that can cause red rot degradation, which is a severe chemical degradation of vegetable-tanned leather due to the used tanning compounds as well as to interactions with atmospheric pollutants [1, 18, 19]. Red-rotted leather have powdery surface, are quite weak and vulnerable to abrasion or tearing. In the more advanced states, the red-rotted leathers became quite red.

Even though the cover of Mus. 243 appeared powdery and fragile, the pH difference was 0.3 indicating no risk of red rot degradation. The powdery and friable appearance of the cover could be mainly ascribable to the usage of the manuscript over time and to the acidic treatment to which the cover had been subjected to obtain a marbled aspect.

A more valuable indicator of the stability and the hydrothermal stability of collagen is the measure of the shrinkage temperature [14] that represents a reliable measure of the degree of deterioration of collagen fibres. For this measurement only very small amount of material is needed.

The triple helix of collagen consists of polypeptidic chains held together by hydrogen bonds and cross-links to form elongated fibrils, which bond together to give rise to fibres with ordered structure. When heated in water, the hydrogen bonds break and collagen deforms to randomly disordered chains in a specific temperature interval. The deformations appear as a shrinkage of the fibres, and the temperature interval where the shrinkage takes place is a measure of the physical stability of collagen.

The shrinkage activity of collagen can be divided into five different intervals with different characteristics of the activity and temperatures. The intervals are conventionally [14] indicated as:

- A1 to B1: distinct shrinkage activity is observed in individual fibres. TF ($T_{first}$) is the temperature at which the first shrinkage activity is recorded; TB1 is the initial temperature of subsequent B1 activity.
- B1 to C: shrinkage activity of one fibre (occasionally more) immediately followed by shrinkage activity in another fibre is observed in the temperature interval from TB1 to TS, where TS ($T_{shrinkage}$) represents the starting temperature of the main interval C.
- C to B2: is the main shrinkage interval, and the temperature to which this activity begins, is indicated as shrinkage temperature TS. In this interval almost all fibres shrink simultaneously and continuously. The ending temperature of the main shrinkage interval is conventionally indicated as TE ($T_{end}$).
- B2 to A2: shrinkage activity of fibres, which did not contract at lower temperature in the previously described intervals. Fibres do not shrink simultaneously (TE to TA2 interval).
- A2 to the last event: the shrinkage is going to end, but few fibres contract with a well distinct and not simultaneous event. The activity stops at the end of this interval TA2-TL, where TL ($T_{last}$) is the temperature at which the last shrinkage is observed.

The temperature range of the whole observed activity is $\Delta T_{total} = TL - TF$, whereas the length $\Delta T = TE-TS$ of the C phase corresponds to the shrinkage interval. The hydrothermal stability of the original leather was analysed before and after the nanocollagen treatment.

Three measures were performed on the original leather before and after the treatment with the chosen nanocollagen solution. Results are reported in Table 6; Fig. 4 shows the shrinkage activity of some fibres not treated and treated with nanocollagen.

*Table 6*
*Shrinkage temperatures (°C) of the original leather of Mus. 243 cover, before and after the treatment with nanocollagen 0.7 mg/ml in isopropyl alcohol/water 70/30 (V/V). Average on 3 samples for each treatment.*

| Sample | Initial temperature of shrinkage intervals | | | | | Final shrinkage temperature | Shrinkage temperature total interval |
|---|---|---|---|---|---|---|---|
| | TF (A1) | TB1 (B1) | TS (C) | TE (B2) | TA2 (A2) | TL | $\Delta T_{total}$ |
| non-treated | 33.1 ± 0.6 | 41.1 ± 1.3 | 47.6 ± 1.3 | 70.3 ± 1.1 | 80.6 ± 4.1 | 93.5 ± 3.6 | 60.4 ± 3.0 |
| treated | 37.2 ± 3.7 | 41.6 ± 5.2 | 59.6 ± 4.0 | 76.8 ± 0.0 | 77.8 ± 0.5 | 79.9 ± 0.8 | 42.7 ± 2.9 |

TF=$T_{first}$; TS=$T_{shrinkage}$; TE=$T_{end}$; TL=$T_{last}$

The leather treated with the nanocollagen solution showed a significant increase in the shrinkage temperature: 47.6°C before the treatment and 59.6°C after nanocollagen application. The shrinkage interval $\Delta T$ (59.6 - 76.8°C) is shorter for the treated leather, indicating an increased uniformity in the fibres length and arrangement, but occurs at higher

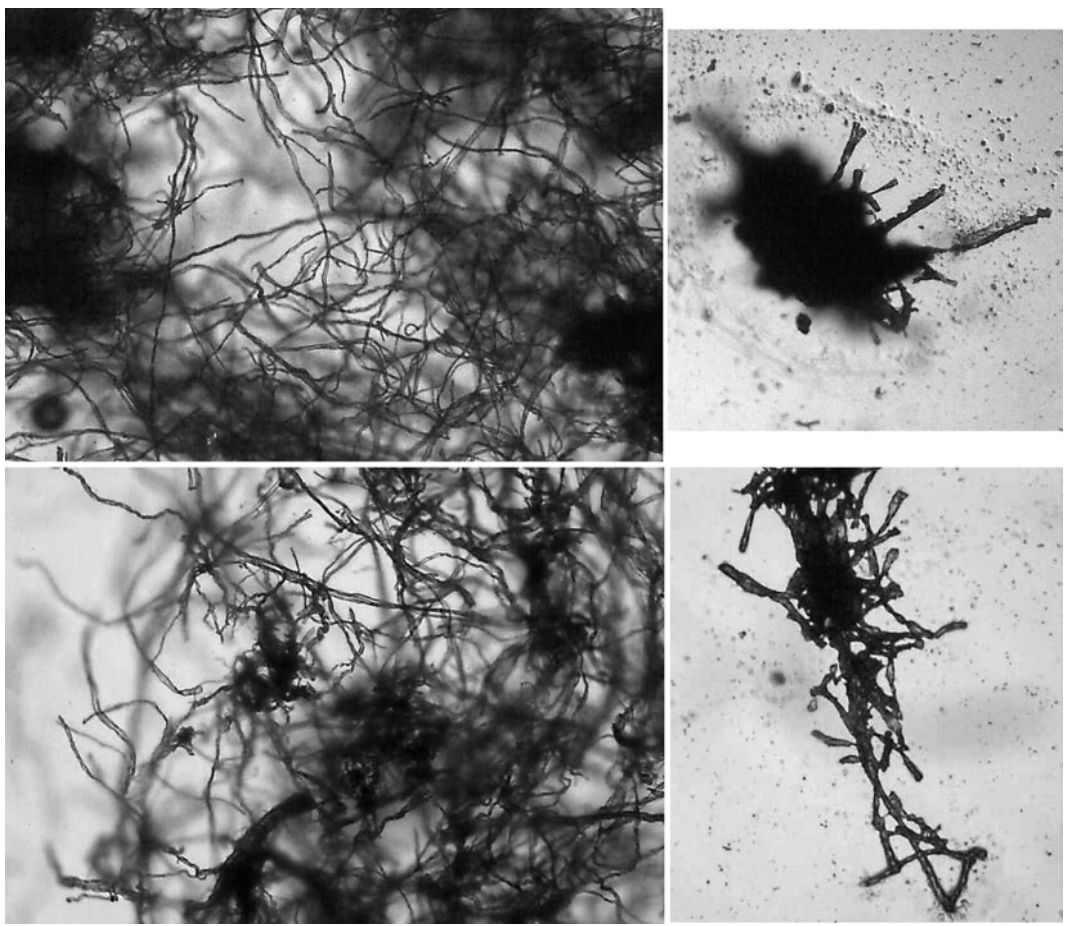

**Fig. 4.** Shrinkage activity of the original Mus. 243 leather. Top: non-treated sample before shrinkage (left) and at the end of the shrinkage activity (right). Bottom: sample treated with nanocollagen solution 0.7 mg/ml in isopropyl alcohol/water 70/30 (V/V) before shrinkage (left) and at the end of the shrinkage activity (right). Magnification 50X. At the end of the shrinkage activity, the non-treated fibres are fragmented and collapsed, while the treated partially maintain their integrity.

temperatures, highlighting the reconstitution of the bulk of collagen fibres.

It is to underline that undamaged standard collagen usually shrinks at TS around 65°C [20] not far from the values reached by the original leather after the treatment with nanocollagen.

At the end of all the experimental work, the nanocollagen solution in isopropyl alcohol/water 70/30 (V/V) was applied by spraying to the flesh side of the Mus. 243 cover.

Due to the high brittleness of the cover leather, to ensure a more effective consolidation, it was decided to apply the solution also to the hair side, but, to avoid an excessive increase in stiffness, the solution was applied at reduced concentration (0.35 mg/ml). Moreover nanocollagen, due its capability to link fibres together, was used to adhere small fragments of the cover, which were partially lifted or detached (Fig. 5).

## 5. Conclusions

The research presented in this paper allowed for the synthesis of a new kind of nanostructured nanotubes of collagen especially conceived for the preservation of leather artefacts, and able to ensure a real and durable effectiveness of the treatment, without altering the appearance and the equilibrium of the internal fats and moisture of the leather.

The nanomaterial was developed to exhibit novel characteristics, such as increased strength, flexibility and solubility in unusual solvents.

The mechanical tests demonstrated a very positive increase in tensile and tearing resistance, as well as in bending resistance, high enough to contrast the weakness of the original artefact, without creating a rigid and breakable leather.

SEM imaging well evidenced the capability of the nanocollagen solution to create bonds between the collagen fibres and to fill the inter-fibres cavities. Moreover, after nanocollagen application, the fibres appeared less tangled than before and the leather surface appeared smoother and rebuilt in its integrity. The measurements of shrinkage temperature showed an improvement in the response of collagen fibres to the degradation, with a partial reconstitution of the lost elasticity and flexibility. This behaviour can be explained as a rehydration and a partial restoration of the triple helixes bonds and the cross-links between the superhelices forming the quaternary structure of the collagen molecule.

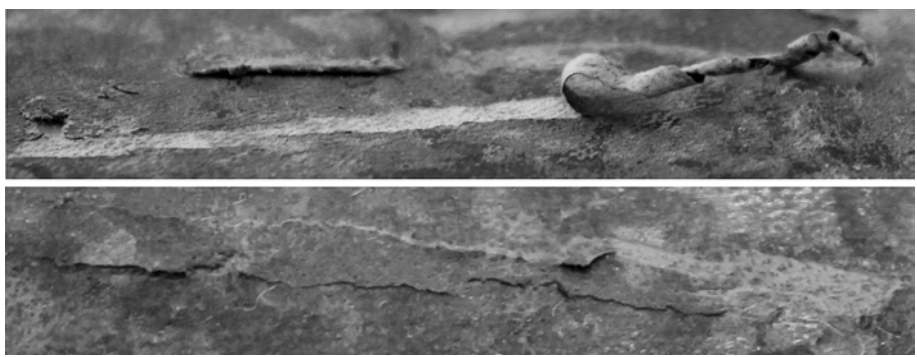

**Fig. 5.** Top: Area of the leather cover presenting partial detachment and lifting. Bottom: same area after the nanocollagen application.

The aforementioned effects made the new synthesised nanocollagen particularly suitable for the treatment of the volume Mus. 243, which complete restoration is described in a Master Thesis for the high training school (SAF) of the Istituto centrale restauro e conservazione patrimonio archivistico e librario [21].

Further tests of the developed product were performed by applying it to a very deteriorated leather armchair, with excellent results.

The mass production and marketing of the novel product is under study.

**Appendix A. Supplementary data**

## Supplementary data

*Confocal Scanning Laser Microscopy (CLSM).*

Leather samples were analysed with confocal laser scanning microscope (CLSM), FV1000 (Olympus Corp., Tokyo, Japan), using laser channels at λ excitation 488 nm and emission range 459-494 nm; and at λ excitation 635 nm and emission range 606-686 nm.

CLSM data are a set of two-dimensional (2D) cross-sectional images in the x-y plane obtained with IMARIS 6.2 software (Bitplane AG, Zurich, Switzerland). A set of three-dimensional (3D) values and cross-sectional images in the xz-yz plane, obtained with IMARIS 6.2 software, were also recorded.

Fig. SI1 displays two images of the flesh side of leather non-treated and after application of nanocollagen at concentration 0.7 mg/ml in isopropyl alcohol/water 70/30 (V/V).

By comparison between the two images the filling and reconstructing effect of nanocollagen is quite evident; furthermore, after application of the product, the leather surface appears smoother than before.

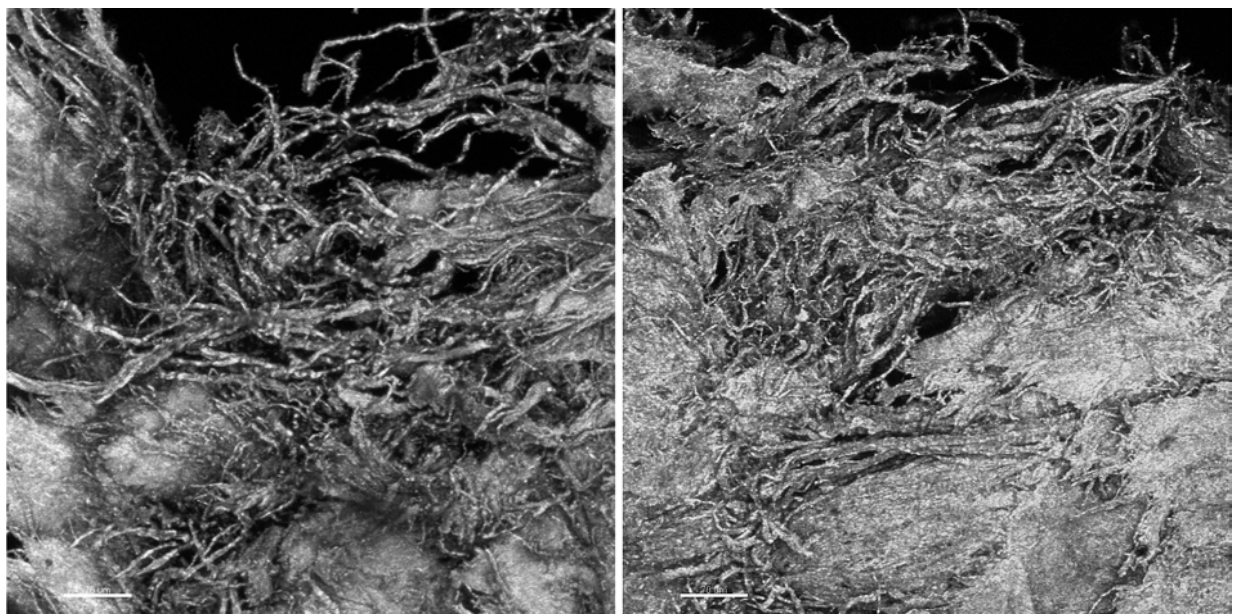

**Fig SI1**: CLSM images. Left: non-treated leather. Right: the same leather after application of nanocollagen. Magnification 20X. Laser excitation: 635 nm.

*Attribution of the characteristics bands of nanocollagen by FTIR.*

Table S1. Infrared spectral absorption bands for the nanocollagen final products.

| Sample | Wavenumber (cm$^{-1}$) | Functional group | Reference |
|---|---|---|---|
| **Nanocollagen** | 3316 | N-H stretching | *Spectrometric Identification of organic compounds, Seventh Edition, Robert M. Silverstein, Francis X. Webster David J. Kiemle* |
| | 1635 | C(=O) Amide I stretching | |
| | 1548 | N-H Amide II bending | |
| | 1454 | C-N stretching | |
| | 1237 | Interaction between N-H bending and C-N stretching | |